\documentclass[12pt,preprint]{aastex}

\shorttitle{An ultracompact X-ray binary in NGC 1851}
\shortauthors{Zurek et al.}
\def\simlt{\mathrel{\rlap{\lower 3pt\hbox{$\sim$}}
        \raise 2.0pt\hbox{$<$}}}
\def\simgt{\mathrel{\rlap{\lower 3pt\hbox{$\sim$}}
        \raise 2.0pt\hbox{$>$}}}

\begin{document}

\title{An ultracompact X-ray binary in the globular cluster NGC
1851$^1$} \footnotetext[1]{Based on observations with the NASA/ESA
Hubble Space Telescope, obtained at the Space Telescope Science
Institute, which is operated by the Association of Universities for
Research in Astronomy, Inc. under NASA contract No. NAS5-26555.}

\author{D.~R.~Zurek$^2$, C.~Knigge$^3$, T.~J. Maccarone$^3$,
A.~Dieball$^3$ and K.~S.~Long$^4$} \footnotetext[2]{Department of
Astrophysics, American Museum of Natural History, New York, NY 10024}
\footnotetext[3]{School of Physics and Astronomy, University of
Southampton, SO17 1BJ, UK} \footnotetext[4]{Space Telescope Science
Institute, Baltimore, MD 21218}

\begin{abstract}

We present far-ultraviolet photometry obtained with the {\em Hubble Space Telescope} of the low-mass X-ray binary 4U 0513-40 in the globular cluster NGC~1851. Our observations reveal a clear, roughly sinusoidal periodic signal with $P \simeq 17$~min and amplitude 3\%-10\%. The signal appears fully coherent and can be modelled as a simple reprocessing effect associated with the changing projected area presented by the irradiated face of a white dwarf donor star in the system. All of these properties suggest that the signal we have detected is orbital in nature, thus confirming 4U 0513-40 as an ultracompact X-ray binary (UCXB). All four confirmed UCXBs in globular clusters have orbital periods below 30 minutes, whereas almost all UCXBs in the Galactic field have orbital periods longer than this. This suggests that the dynamical formation processes dominate UCXB production in clusters, producing a different orbital period distribution than observed among field UCXBs. Based on the likely system parameters, we show that 4U 0513-40 should be a strong gravitational wave source and may be detectable by LISA over the course of a multi-year mission.

%Based on the likely system parameters, we show that 4U 0513-40 should be a strong gravitational wave source and detectable by LISA. 

%All four confirmed UCXBs in globular clusters have orbital periods below 30~minutes, wheras all five confirmed UCXBs in the Galactic field have orbital periods longer than this.

\end{abstract}

\keywords{globular clusters: individual(\objectname{NGC 1851}) --
stars: close binaries -- stars: individual {4U 0513-40} --
ultraviolet: stars}

\section{Introduction}

It has been known for several decades that bright low-mass X-ray
binaries (LMXBs) are $\sim 100$ times overabundant in Globular
clusters (GCs) relative to the Galactic field (Katz 1975; Clark
1975). More specifically, GCs contain 13 of the $\sim 100$ bright
Galactic LMXBs, but only $\sim$0.01\% of the total stellar mass
content of the Galaxy. The reason for this is thought to be the
existence of {\em dynamical} LMXB formation channels, which are only
available in dense GC cores. Potential channels include the direct
collision of a neutron star (NS) with a red giants (Verbunt 1987,
Davies et al.~1992, Ivanova et al.~2005), the tidal capture of a main
sequence star by a NS (Fabian, Pringle \& Rees 1975; Bailyn \&
Grindlay 1987) and exchange interactions between NSs and primordial
binaries (Hilles 1976; Rasio et al.~2000).

If the dominant LMXB formation channels are different in GCs and the
Galactic field, the properties of their respective LMXB populations
may also be different. In particular, most
of the bright LMXBs in GCs might be ultracompact X-ray binaries
(UCXBs; Bildsten \& Deloye 2004, Ivanova et al.\ 2005). UCXBs, which
are interacting binaries with extremely small binary separations ($a
\simeq 10^{10}$~cm) and short orbital periods ($P_{orb} \simlt$ 1 hr),
appear to be rare amongst the Galactic field LMXB population: the list
of confirmed UCXBs (with measured $P_{orb}$) in in't Zand et
al. (2007) contains only 5 objects that belong to this population. By
contrast, 3 of the 13 GC LMXBs are confirmed UCXBs; these are
4U\,1820-30 in NGC\,6624 ($P_{\rm{orb}} = 11.4$ min, Stella et
al.~1987), 4U\,1850-087 in NGC\,6712 ($P_{\rm{orb}} = 20.6$ min, Homer
et al.~1996) and CXO\,J212958.1+121002 in M\,15 (=M15-X2; Dieball et
al. 2005), with several more suggested to be ultracompact X-ray
binaries on the basis of more indirect evidence (see e.g. Verbunt \&
Lewin 2006 for a review).

Since the period distribution of GC LMXBs may be a direct tracer of
the dynamical close encounters taking place in GC cores, it is
important to establish orbital periods for as many of these sources as
possible. Doing so could also lead to a significant increase in the
size of the total UCXB sample. This is desirable, because UCXBs are
astrophysically important systems in their own right. This is
because they are laboratories for accretion and binary evolution in
extreme settings, and because they are strong gravitational
wave sources that may be detectable by LISA (Nelemans \& Jonker 2006;
Nelemans 2009). 

Here, we present time-resolved, far-UV photometry of the LMXB 4U
0513-40 in NGC 1851, which was suspected to be a UCXB based on several 
strands of circumstantial evidence (Deutsch etal  2000; Verbunt 2005;
Nelemans \& Jonker 2006; in't Zand etal. 2007). Our far-UV data of
this system contain a $\simeq$~17~min periodic signal that is present
in all four observing epochs, is consistent with being coherent and is
probably caused by a reflection effect associated with the irradiated
surface of the donor star in this system. Based on all this, we argue
that the observed periodic variability is an orbital signature, and thus
that 4U 0513-40 should be regarded as a confirmed UCXB with
$P_{orb} = 17$~min. 

\section{Observations and data reduction}
\label{data}

NGC 1851 was observed three times with the F140LP filter in the Solar
Blind Channel (SBC) of the Advanced Camera for Surveys (ACS) on board
the {\it HST}. This instrument/detector/filter combination has a plate
scale of 0.032\arcsec~pixel$^{-1}$, a pivot wavelength of $\lambda_p =
1527$~\AA, and an rms bandwidth of $\Delta_{\lambda} = 125$~\AA. All
of the observations took place in August of 2006. Each observing
epoch consisted of 4 {\it HST} orbits, broken up into a series of 90
second exposures. In total, we obtained 273 of these exposures. In
addition, we also examined archival data taken in March of 1999 with
the Space Telescope Imaging Spectrograph (STIS), using the
FUV-MAMA/F25QTZ detector/filter combination, with a plate scale of
0.025\arcsec~pixel$^{-1}$, $\lambda_p = 1595$~\AA and
$\Delta_{\lambda} = 97$~\AA.

A full description of the data, as well as their reduction and
analysis will be provided in a separate publication (Zurek et
al. 2009, in preparation). Briefly, all of the FUV count rates and
magnitudes presented in this paper were calculated via standard aperture
photometry techniques, as implemented in the {\em daophot} package
within {\it IRAF}\footnote{{\tt IRAF}
(Image Reduction and Analysis Facility) is distributed by the National
Astronomy and Optical Observatories, which are operated by AURA, Inc.,
under cooperative agreement with the National Science
Foundation.}. For the photometry on our ACS/SBC (STIS/FUV-MAMA)
images, we used an aperture radius of 4 (7) pixels and a sky annulus
extending from 10 to 20 (15 to 35) pixels. Aperture photometry is
sufficient for our purposes because the FUV image is not particularly
crowded (see Figure~1). 

The wavelength-dependent throughput curves of the ACS/SBC/F140LP and
STIS/FUV-MAMA/F25QTZ instrument/detector/filter combinations are
very similar, though not identical. Therefore we checked
for far-UV variability by comparing the ACS and STIS count rates,
after correcting for throughput differences and the different
photometric aperture sizes and background regions that were used. We
have calculated this correction factor from a set of (mostly
blue horizontal branch) stars that are common to both sets of
images. We find that for these stars, our ACS/SBC count rates are 3.3
times larger than our STIS/F25QTZ ones.

\section{Analysis and Discussion}
\label{analysis}

Homer et al. (2001) have already used the HST/STIS/F25QTZ observations
to identify the optical/far-UV counterpart of 4U 0513-40. They confirm the
suggestion of Deutsch et al. (2000) that ``Star A'' (in the
nomenclature of Deutsch et al.) is the correct counterpart to the
LMXB, while two other blue sources previously suggested as possible
counterparts by Auri\'{e}re, Bonnet-Bidaud \& Koch-Miramond (1994),
designated as X-1 and X-2b, are inconsistent with the precise Chandra
position of 4U 0513-40. Figure~1 shows the location of these 3
sources in our ACS/SBC images. 

Since Homer et al. (2001) had already reported a negative search for
far-UV variability associated with Star A in the STIS observations, we
started by focusing on the newer, higher signal-to-noise ACS data. The
ACS-based far-UV light curves of all 3 objects are shown in
Figure~2. It is immediately obvious that Star A, the counterpart
proposed by Deutsch et al. (2000) and Homer et al. (2001), does, in
fact, exhibit strong far-UV variability, especially between the 3
observing epochs. For example, the mean count rate drops by a factor
of about 2 between epochs 2 and 3. No similar change in count rate is
seen in either X-1 or X-2b, which bracket Star A in far-UV brightness.

We then searched for {\em periodic} signals in the data by carrying
out a power spectral analysis. Figure~3 shows the Lomb-Scargle power
spectra calculated for all 3 sources from the combined ACS data
sets. It is immediately obvious that only Star~A shows clear evidence
of a signal at a frequency other than HST's orbital frequency
($f_{HST}$). More specifically, there is an obvious peak at a
frequency of about $f_{orb} \simeq 85$~c~d$^{-1}$ (which corresponds
to a period of $P_{orb} \simeq 17$~min).

In an effort to better understand the Star~A power spectrum
(specifically the power excess at $f_{HST}$ and the sidebands at $f_{orb}
\pm f_{HST}$), we  have created a simple model light curve. This model
has exactly the same time sampling as the observations, contains 
a single 17-min periodic signal with amplitude 0.75~c~s$^{-1}$ and
accounts for the dominant long-term trend in the data by setting the
simulated mean count rate in each epoch equal to the observed one. The
power spectrum generated from this simple noise free model is shown in
the second panel of Figure~3 (labelled ``Star~A Simulation''), and
clearly captures all of the main features of the data. This shows that
both the sidebands around $f_{orb}$ and the strong apparent signal at
$f_{HST}$ are sampling artifacts. The latter, in particular, is due to 
leakage from the low-frequency power excess associated with the
long-term variability. 

In order to test if the 17~min signal is persistent, we carried out
power spectral analysis on each epoch independently, where we now also
included the STIS data that was obtained 7 years before the ACS
observations. The result is shown in Figure~4. The periodicity was
indeed present in all 4 epochs (1 STIS + 3 ACS), but was noticeably
weaker in the STIS data and in the third ACS epoch. Considered in
isolation, the power excess at $f_{orb}$ in each of those epochs would
be marginally significant at best.

Both Figures~3 and 4 show that there are several possible aliases
associated with the 17~min signal. In order to establish the relative
likelihoods for each of these and get estimates of the errors
associated with each of them, we carried out a bootstrap
analysis. Only the ACS data was used for this purpose, since the
signal is weakest in the STIS data and since no unique cycle count can
be assigned to the large time interval between the STIS and ACS
observations. We began by removing long-term trends from the data
stream by estimating the mean count rate in each HST orbit and
subtracting these averages from the data. Next, we created 1000 mock
time series by sampling with replacement from the orbit-mean-subtracted
ACS data. Finally, we created power spectra for all of these fake data 
sets and recorded the frequencies corresponding to peak flux. The
histogram created from this set of peak frequencies is shown in
Figure~5 (superposed on the orbit-mean subtracted power spectrum for
all of the data, which is shown as the shaded region). As expected from
the power spectrum, there are clearly four dominant plausible
aliases. The relative number of bootstrap trials associated with each
alias is a measure of the likelihood that this alias is the correct
one (see Southworth et al 2006, 2007, 2008 and Dillon et al. 2008). 
Similarly, the location and width of the histogram peak
associated with each alias provides a simple estimate of the period
and period error for this alias. Our final results for the four viable
aliases are:

\begin{eqnarray}
P_1 = 16.8078 \pm 0.0013 \; \; \; ({\rm Prob} = 42.2\%) \\
P_2 = 16.9122 \pm 0.0017 \; \; \; ({\rm Prob} = 35.8\%) \\
P_3 = 16.9362 \pm 0.0016 \; \; \; ({\rm Prob} = 13.9\%) \\
P_4 = 16.8678 \pm 0.0559 \; \; \; ({\rm Prob} =  8.1\%)
\end{eqnarray}

We also tried to establish more directly whether the periodic
signal is consistent with being coherent across all of our epochs and
whether the evidence for a changing amplitude is compelling. To
this end, we carried out a least-squares fit to the orbit-mean-subtracted
STIS+ACS data set for Star~A. In this fit, the period was kept fixed
at $P_1$\footnote{In fact, the period was fixed to be one of the many 
sub-aliases of $P_1$ that are present in the power spectrum of the
{\em combined} STIS+ACS data.}, so the free parameters were the phase
and amplitude (both assumed to be constant across the data). The
result of this fit is shown in Figure~6. It illustrates rather clearly
that the amplitude is not constant across epochs. As already noted
above, the signal is strongest in ACS epochs 1 and 2 and weakest in
the STIS data and ACS epoch 3. The fit also shows that the assumption
of constant phase is reasonable, i.e. the observed signal is
consistent with being coherent across the entire data set.

In order both to test further the coherence properties of the signal,
and to determine its average waveform, we have folded the entire
STIS+ACS data stream for Star~A onto the same (sub-)alias that was
adopted for the fit in Figure~6. The result is shown in Figure~7,
along with the phase-binned average waveform. Again, the data seem to
be consistent with the idea that the signal is fully
coherent. Moreover, the average waveform is fairly simple and roughly
sinusoidal.

As a final step, we carried out separate fits with fixed period to each
individual epoch in order to quantify the changing amplitude of the
17-min signal. Table~1 lists the mean count rates and
absolute as well as fractional amplitudes of the signal as estimated
from these fits. The fractional amplitude varies by over a factor of
3 between epochs, from about 3\% to 10\%, with the largest change
happening across the 6-day gap between ACS epochs 2 and 3. It is also
worth noting that there appears to be a correlation between
the FUV brightness of the system (i.e. the mean count rate) and the
amplitude of the 17-min signal, in the sense that the amplitude is
highest when the system is brightest. This correlation holds
regardless of whether the amplitude is expressed in absolute or
fractional terms. 

It is worth noting that there is no inconsistency between our claim
that the 17-min signal is present in the STIS observations, and the
non-detection of a periodicity in the same data set by Homer et
al. (2001). As mentioned above, the signal is only marginally
detectable if this data set is analysed in isolation. What makes
the weak power excess in the STIS observations convincing is that it
is located at the same frequency as the obvious signal in the ACS data
set. Indeed, Homer et al. (2001) place a formal upper limit of 5\% on
the amplitude of any signal between 5~min and 6~hrs. This is
entirely consistent with our own estimate of a 4\% amplitude for the
17~min signal in the STIS data (see Table~1). 

\section{Discussion}
\label{discussion}

We have shown that the FUV counterpart to 4U~0513-40 in NGC~1851
exhibits a clear periodic signal with $P \simeq 17$~min. This signal
is roughly sinusoidal, has an amplitude of 3\%-10\%, is present in all
four FUV observing epochs and is consistent with being fully
coherent. These properties are in line with those of the orbital
signals seen in UV/optical observations of other UCXB (e.g. Homer et
al. 1996; Anderson et al. 1997; Dieball et al. 2005) and with a simple
model in which these signals are due to a ``reflection effect''
associated with the irradiated donor star in the system (Arons \& King
1993).\footnote{The term ``reflection effect'' is actually somewhat
misleading, since the UV/optical light from the irradiated face is due
to reprocessing, not reflection. Our use of the term here reflects
standard usage in the close binary literature.}

In the Arons \& King (1993) model, there are two components that contribute to the FUV/optical light: the irradiated front face of the donor star (which is likely to be a low-mass, Helium-core white dwarf), and the accretion disk, whose energy budget is also dominated by irradiation. The observed orbital signal is due to the changing projected area of the irradiated face of the donor star, with the fractional amplitude being set by the relative contributions of the disk and donor to the phase-averaged FUV/optical light. We have checked that this simple irradiation model matches roughly the fractional FUV signal using reasonable choices of parameters (including a moderate-to-high inclination). The dependence on total FUV brightness is explained straightforwardly by different FUV responses of the donor and disk to variations in the irradiating X-ray luminosity. In this context, a factor of $\simeq$~2 change in total FUV and fractional amplitude would requires roughly an order of magnitude variation in $L_X$. This is consistent with the x-ray  variability of the source in the RXTE data base, where factor of 10 changes in count rate are seen on time-scales as short as weeks. For comparison, the largest previously reported variations in X-ray luminosity from this source were a factor of 5 (Grindlay \& Hertz 1983). It seems likely that additional effects such as changes in radius, changes in reprocessing efficiency, and disk shielding, play a significant role in the variations of the amplitude of the UV oscillations.

%This is large, but not unreasonable, especially given that other effects -- such as changes in disk radius, reprocessing efficiency and donor shielding by the disk -- may also play a role. 

Based on all this, we are confident that the FUV signal we have
discovered is orbital in nature and probably due to a simple 
reflection effect. We therefore confirm 4U~0513-40 as a 
UCXB with $P_{orb} \simeq 17$~min, making it the fourth confirmed UCXB
in a Galactic GC. The identification of this system as a UCXB is
entirely consistent with existing circumstantial evidence regarding
its nature. More specifically, the optical brightness,
X-ray spectrum and burst properties of 4U~0513-40 have been known for
some time to point towards a UCXB classification (e.g. Verbunt 2005). 

Given the extremely short orbital period we have measured, it is interesting to ask if the gravitational wave signal produced by this source would be detectable by LISA. The gravitational radiation strain for a circular orbit is given in a convenient form by Nelemans, Yungelson \& Portegies Zwart (2001) as:
\begin{equation}
h \simeq 5 \times 10^{-22} ({M}_{chirp}/M_\odot)^{5/3} (P_{orb}/1 {\rm hr})^{-2/3} (d/1 {\rm kpc})^{-1},
\end{equation}
where ${M}_{chirp} = (Mm)^{3/5}/(M+m)^{1/5}$, $M$ is the mass of the neutron star primary, and $m$ is the mass of the white dwarf donor, and with the power coming out entirely in the second harmonic, for a circular orbit. If we adopt $M = 1.4 M_\odot$, $d = 12.1$~kpc (Harris 1996) and $m = 0.05 M_\odot$ (suggested for this $P_{orb}$ by the mass-radius relations in Deloye \& Bildsten [2003]), we find a strain of $h = 6\times10^{-24}$.  This is at approximately the 1-$\sigma$ level for a one year integration with LISA, suggesting that the system may be detectable if the mission lifetime is several years. In practice, this will, however, also depend on the strength and frequency-dependence of the gravitational-wave background due to double WDs, which may become dominant in the relevant frequency regime ($f_{orb} \simeq 1$~mHz; e.g. Nelemans et al. 2001). An additional challenge for detecting gravitational radiation from this object will be dealing with the possible effects of acceleration of the binary in the gravitational potential of the globular cluster, which could lead to a measurable frequency drift over several years, for which a correction would have to be made in order to keep all the gravitational wave power in a single frequency bin. The fact that the position and period of the source are already known (and the latter can still be substantially improved before LISA starts) will be helpful in this context and also ameliorate the hidden trials problem.

We finally point out that, as more UCXB periods are being determined, it seems increasingly likely that there are significant differences between the period distributions of field and GC UCXBs. All four of the confirmed GC UCXBs have $P_{orb} < 30$~min, whereas all five of the confirmed field UCXBs have $P_{orb} > 30$~min (Nelemans \& Jonker 2006; in't Zand, Jonker \& Markwardt 2007). Clearly, this comparison cannot yet be taken at face value: the present numbers are still too small and selection effects have not been taken into account. Moreover,  there is probably no absolute dividing line between the two orbital period distributions. For example, the field LMXB 4U 1543-624 probably has an orbital period of around 18 minutes (Wang \& Chakrabarty 2004), even though it is not yet included in the list of confirmed UCXBs given by Nelemans \& Jonker (2006) and in't Zand et al. (2007). Nevertheless, the period distributions are in line with the expectation that different UCXB formation channels should dominate in the two different environments. If so, it may be possible to use the period distribution of field UCXBs as a tracer of binary evolution, and that of GC UCXBs as a tracer of stellar dynamics in dense environments.

\acknowledgments

This work was supported by NASA through grant GO-10184 from the Space
Telescope Science Institute, which is operated by AURA, Inc., under
NASA contract NAS5-26555.  We thank Frank Verbunt for pointing out
that all known X-ray binaries with orbital periods less than half an
hour are in globular clusters.

\begin{deluxetable}{lccc}
\tablecolumns{4} \tablewidth{0pc} \tablecaption{Values determined for
each epoch \label{amplitude}} \tablehead{ \colhead{Epoch} &
\colhead{Mean Count Rate} & \colhead{Absolute Amplitude} &
\colhead{Fractional Amplitude}} \startdata STIS & $5.142 \pm 0.035$ &
$0.20 \pm 0.06$ & $3.8 \pm 1.1$ \\ ACS1 & $8.031 \pm 0.048$ & $0.77
\pm 0.07$ & $9.6 \pm 0.9$ \\ ACS2 & $9.057 \pm 0.056$ & $0.98 \pm
0.08$ & $10.8 \pm 0.9$ \\ ACS3 & $6.747 \pm 0.024$ & $0.22 \pm 0.05$ &
$3.2 \pm 0.7$ \\ \enddata
\end{deluxetable}

\begin{figure}
\plotone{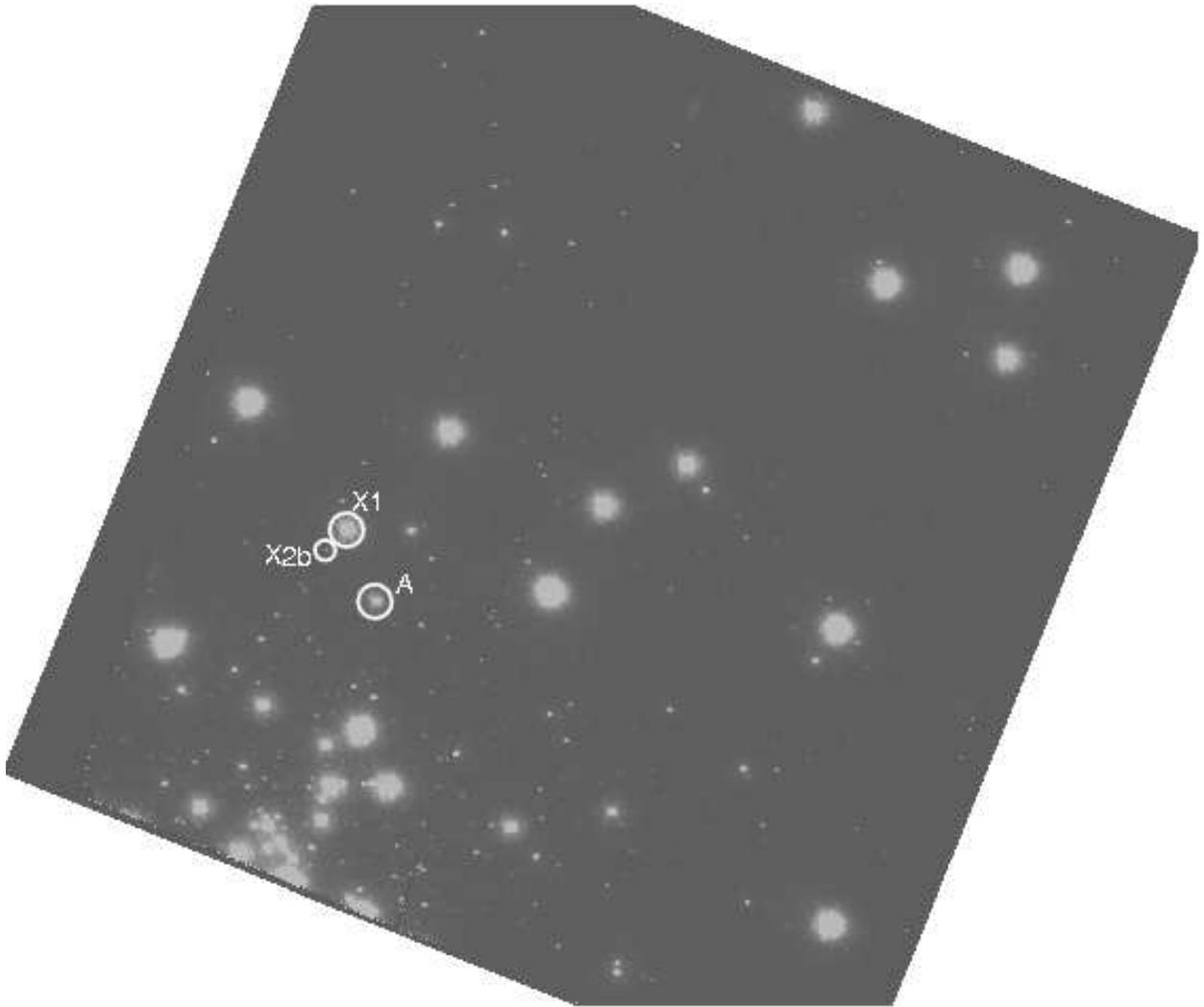}
\caption{The ACS/SBC image with the sources from Homer
identified. North is up and East is left. \label{image}}
\end{figure}

\begin{figure}
\epsscale{0.9}
\plotone{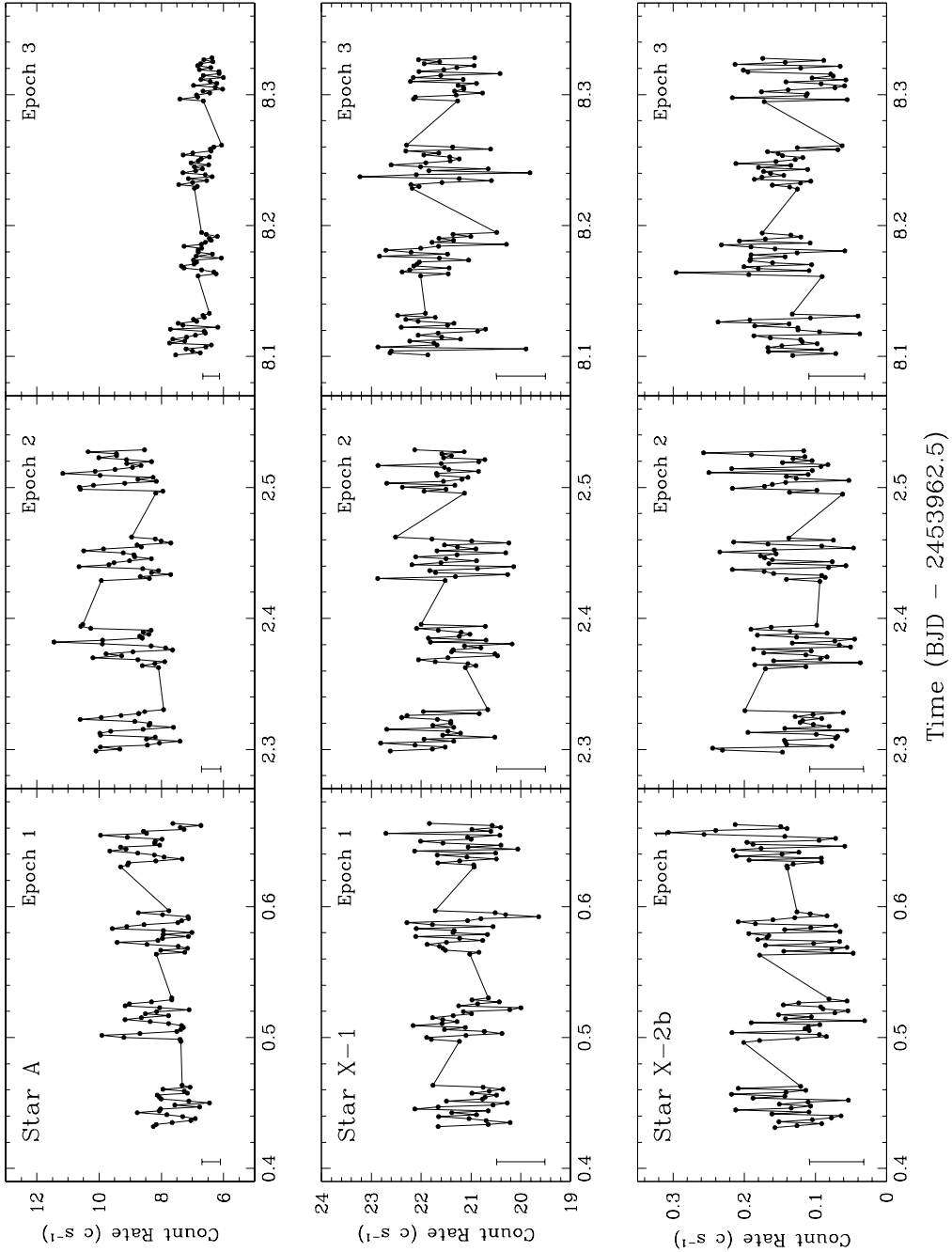}
\caption{ACS light curves for all 3 object that had been suggested as
possible counterparts pre-Homer. Homer et al then showed that Star A
is really the only viable candidate. We have nevertheless shown all
three here to illustrate that Star A is, in fact, by far the most
variable of the lot. The other objects make for convenient "comparison
stars".  \label{light}}
\end{figure}

\begin{figure}
\epsscale{0.85}
\plotone{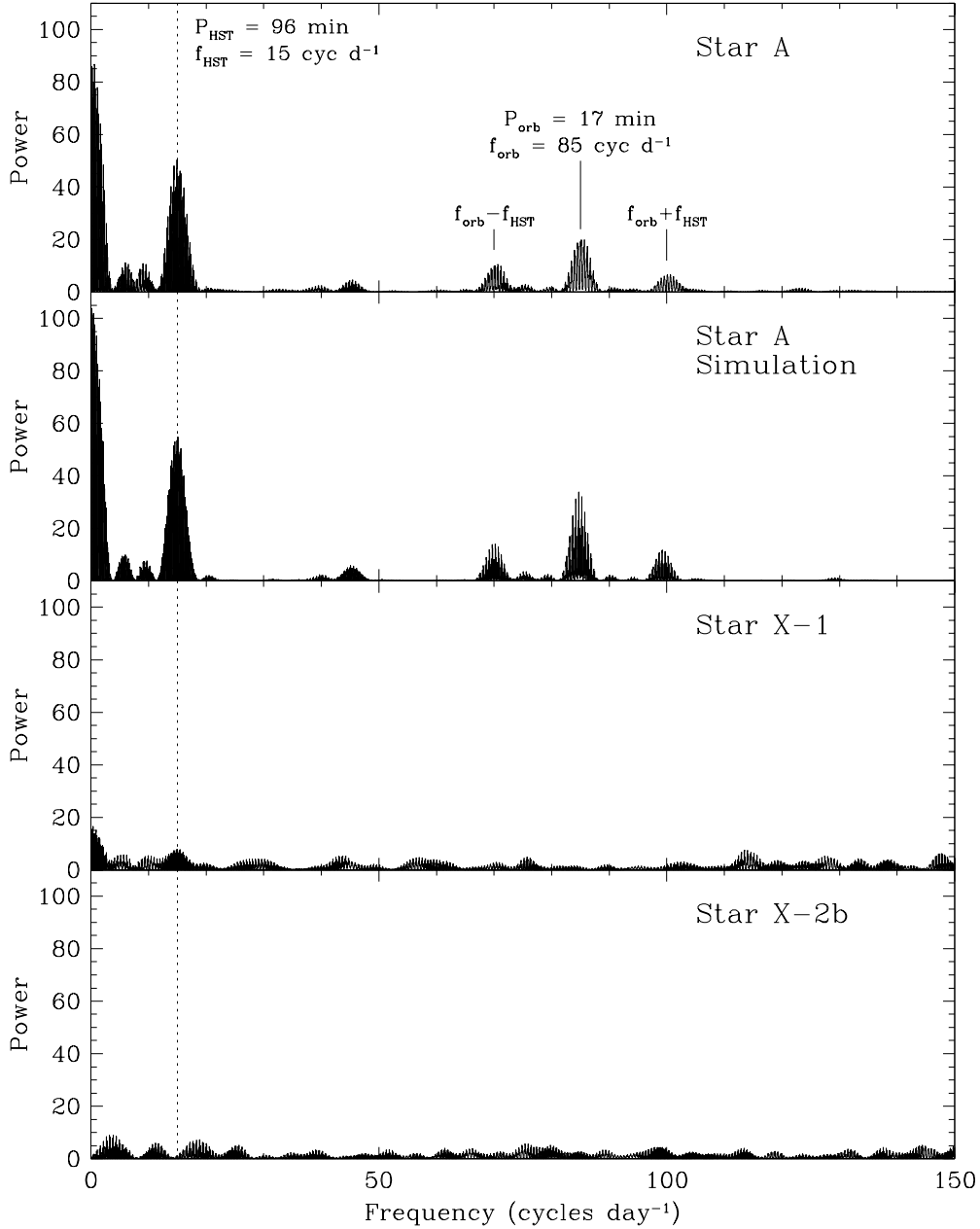}
\caption{Scargle periodograms for all 3 sources. These show clearly
that Star A is the most variable. As expected, the dominant
variability in Star A is due to long-term trends (visible as the peak
rising towards zero frequency).  There is also variability on the HST
orbital period. Those two types of variability are also (more weakly)
seen in Star X-1. However, Star A is the only object to exhibit
another type of obvious signal, namely the $\sim17$ minute
period. This period is indicated, as are its side-bands, which are at
the beat-frequency of the $\sim17$ minute signal and the 96 minute HST
orbital period. So the $\sim17$ minute signal is really the only
"true" signal and is clearly significant. \label{scargle1}}
\end{figure}

\begin{figure}
\epsscale{0.9}
\plotone{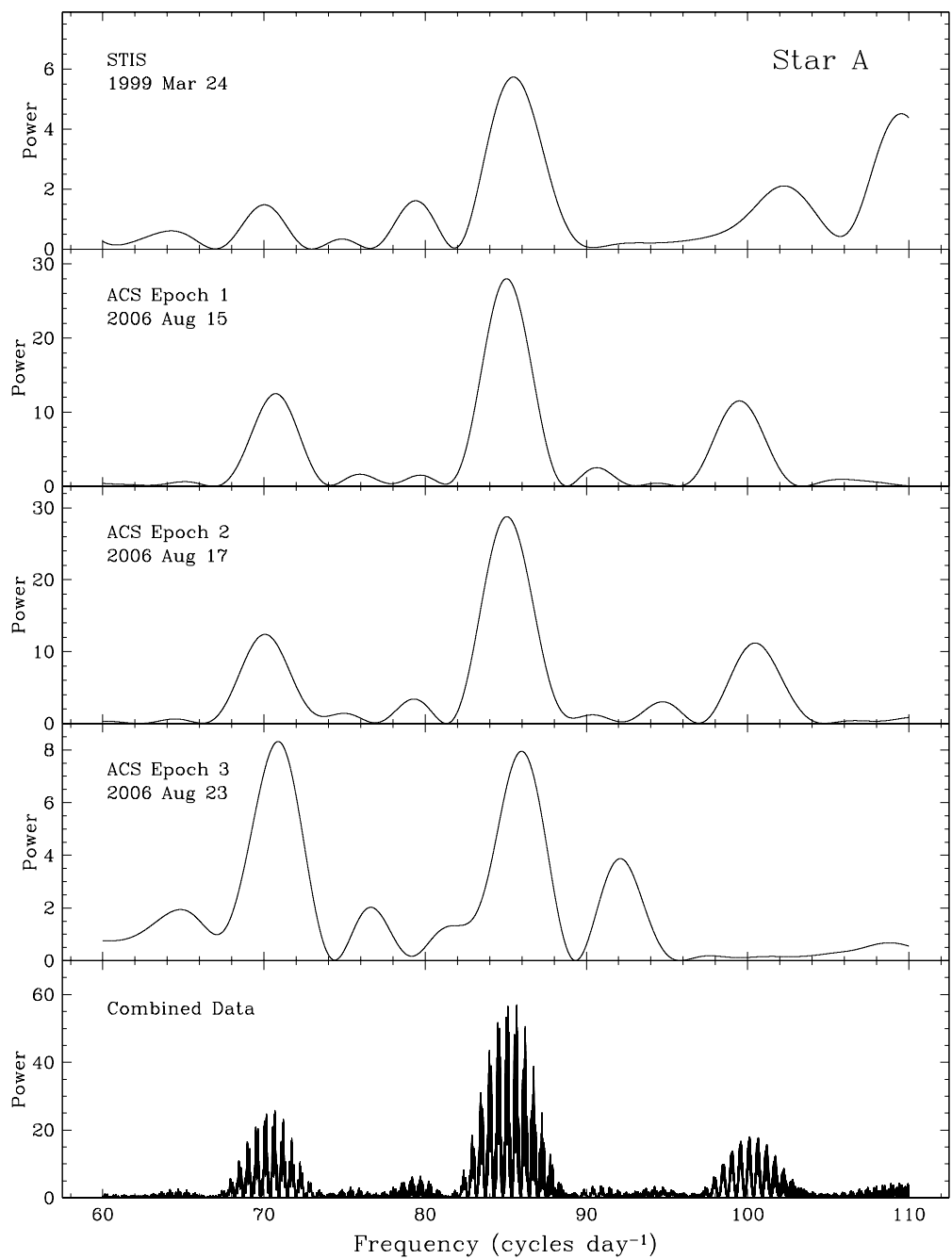}
\caption{A zoom-in on the $\sim17$ minute signal in the individual
epochs (3 ACS and 1 STIS) as well as in the combined data. The signal
is present in all the data sets, however, the signal probably would
not have been considered significant in the STIS data by itself
(typically, scargle powers around 5 are at best marginally
significant). We also note that the ACS epoch 3 has a signal
noticeably smaller than in epochs 1 and 2. This must be a real effect.
\label{scargle2}}
\end{figure}

\begin{figure}
\plotone{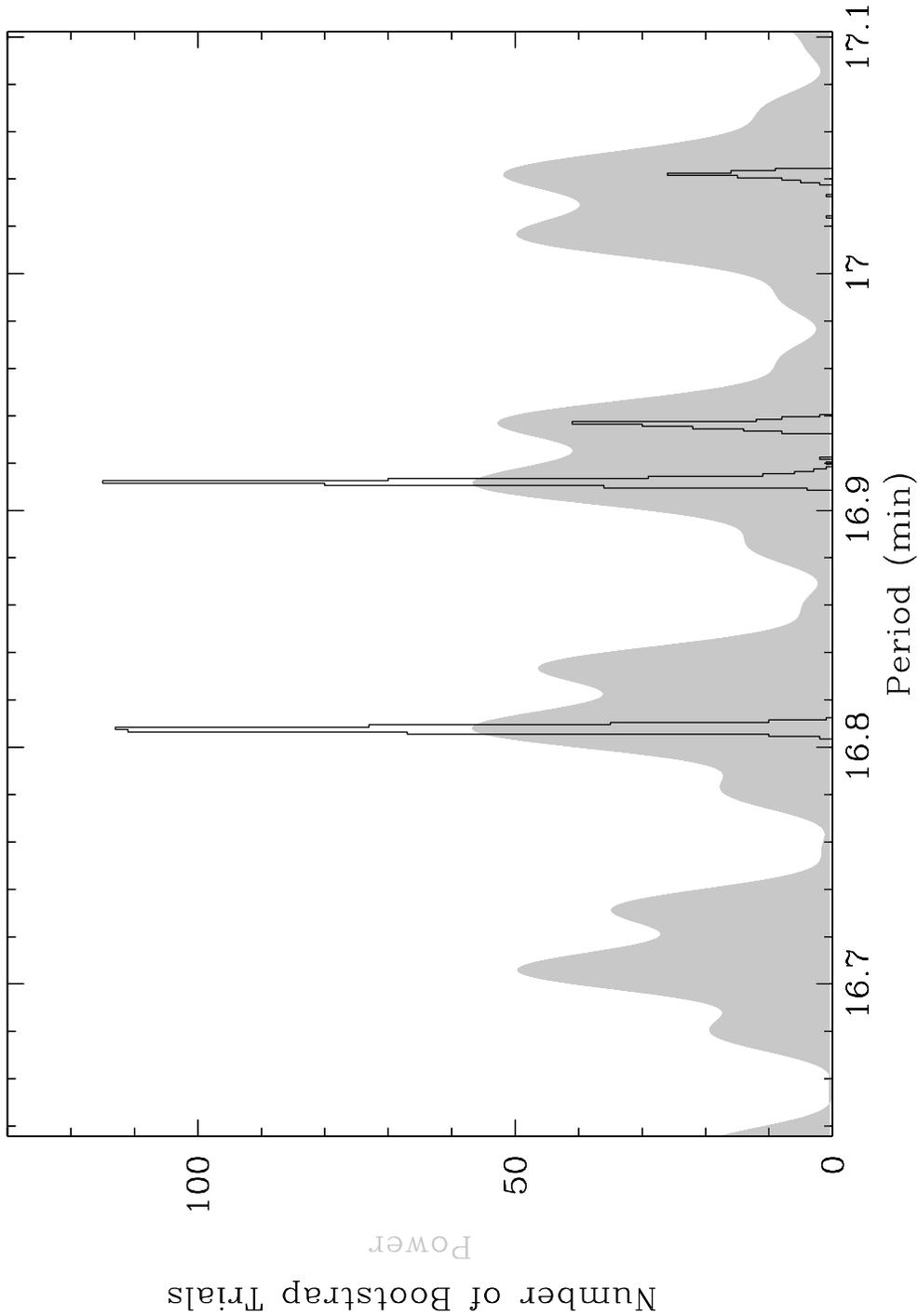}
\caption{The results of the bootstrap analysis of 1000 fake data sets
by sampling with replacement from the orbit-mean-subtracted original
data (ACS+STIS). The power spectra of these fake data sets yield the
histogram shown here (superposed on the orbit-mean subtracted power
spectrum for all the data - shaded region). There are 4 dominant
plausible aliases. \label{boot}}
\end{figure}

\begin{figure}
\plotone{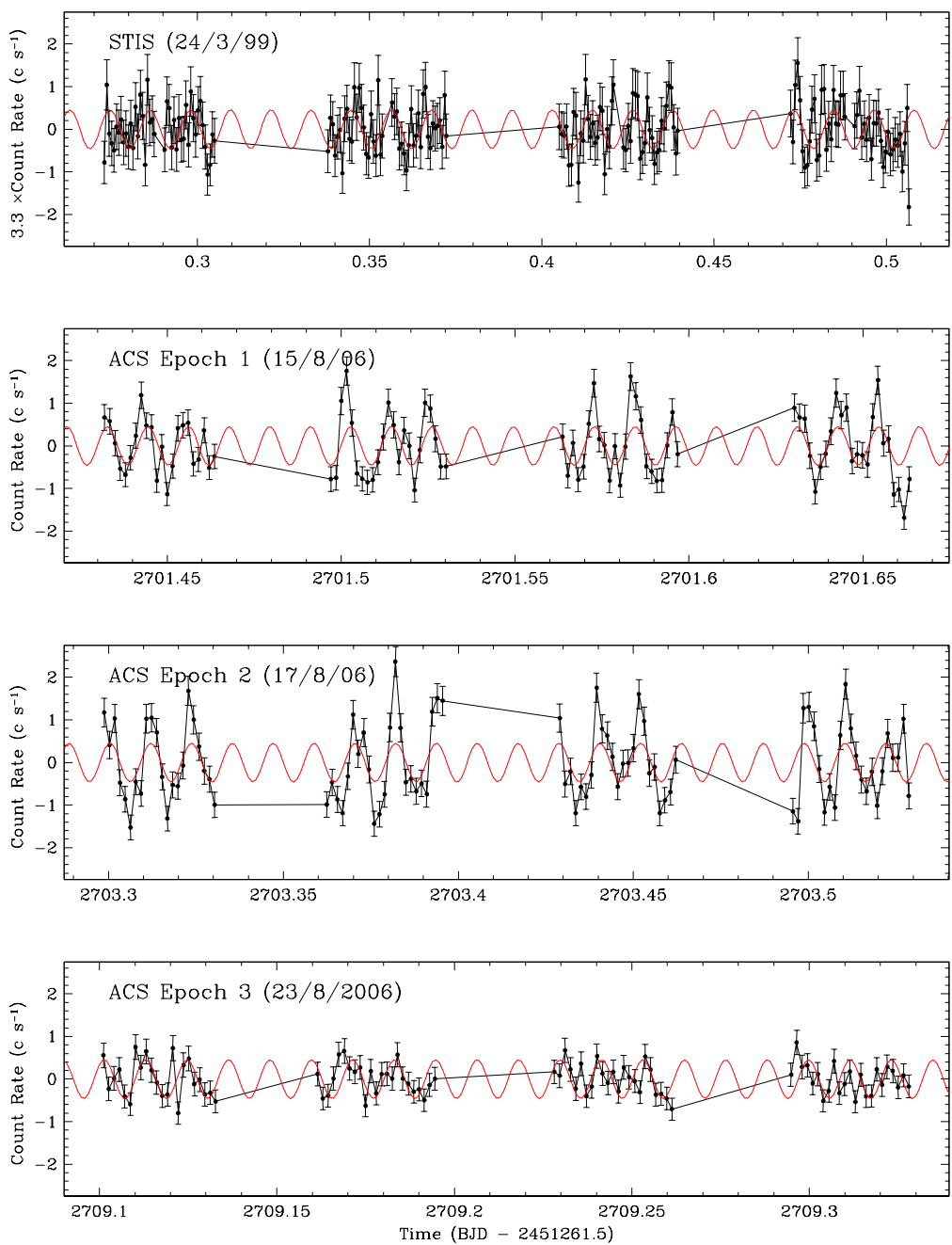}
\caption{We show the orbit-mean subtracted light curves for all the
data (STIS+ACS) along with two models (corresponding to the two
dominant aliases). This demonstrates (a) that there is indeed evidence
for a changing amplitude in the periodic signal and (b) that the data
is at least consistent with being coherent across all of the
data. \label{model}}
\end{figure}

\begin{figure}
\plotone{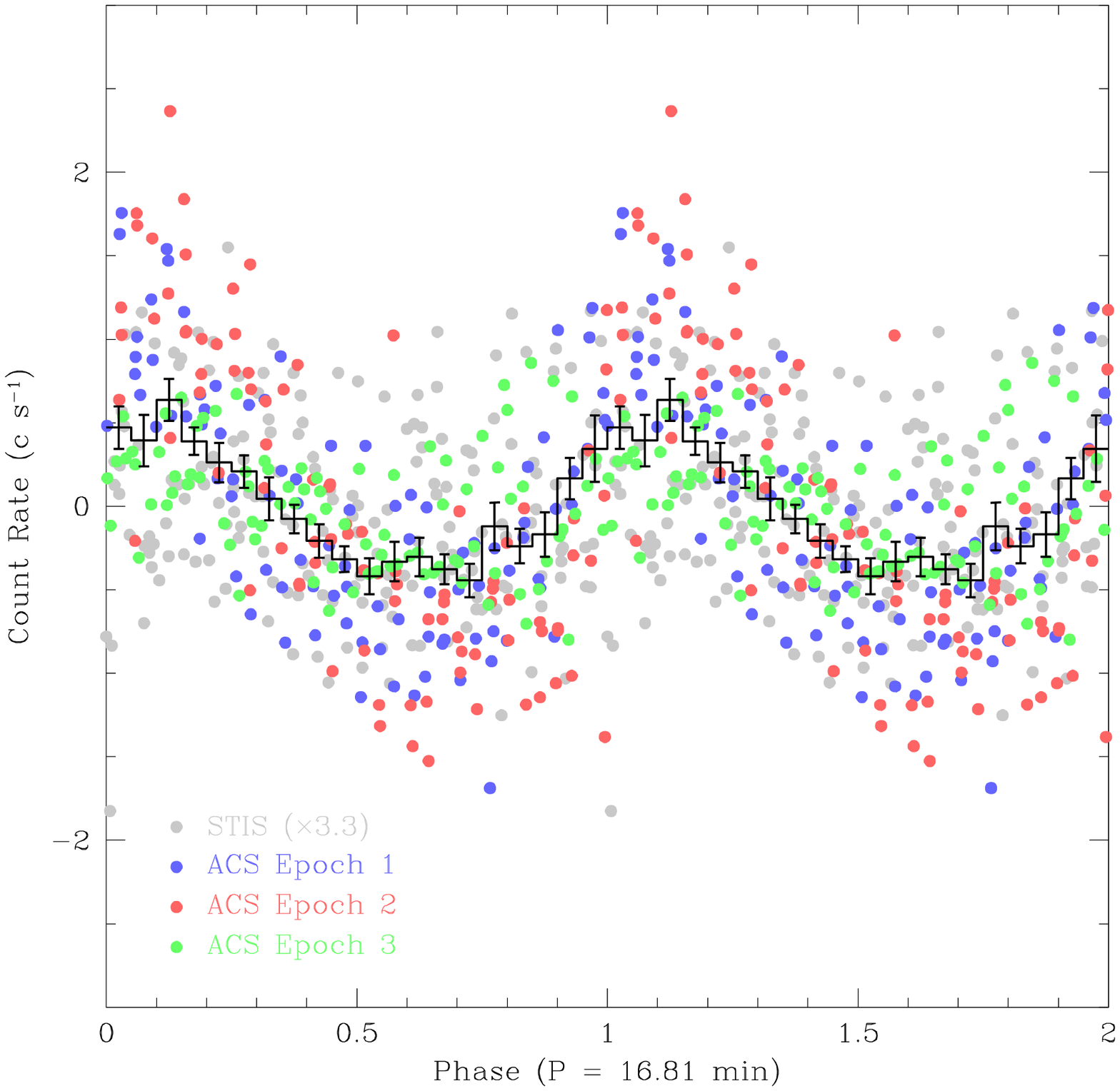}
\caption{The folded signal. The line with error bars is the
phase-folded binned light curve (constructed from the
orbit-mean-subtracted data) and the error are the error of the mean
calculated from all the points in each bin. Each epoch is given a
different color with grey as the STIS epoch, red as ACS epoch 1, blue
as ACS epoch 2, and green as ACS epoch 3. \label{fold}}
\end{figure}


\begin{thebibliography}{}
\bibitem[]{401} Arons, J. \& King, I. R. 1993, ApJ, 413, L121
\bibitem[]{402} Auri{\`e}re, M., Le F{\`e}vre, O. \& Terzan, A. 1984, A\&A, 138, 415
\bibitem[]{403} Bailyn, C. D. \& Grindlay, J. E. 1987, ApJ, 316, L25
\bibitem[]{404} Bildsten, L. \& Deloye, C. J. 2004, ApJ, 607, L119
%\bibitem[]{405} Boroson, B. et al. 2000, ApJ, 545, 399
%\bibitem[]{407} Chakrabarty, D., Homer, L., Charles, P. A. \&
%O'Donoghue, D. 2001, ApJ, 562, 985
%\bibitem[]{} Charles, P. A., Jones, D. C. \& Naylor, T. 1986, Nature
%323, 417
\bibitem[]{408} Clark, G. W. 1975, ApJ, 199, L143
\bibitem[]{409} Davies, M. B., Benz, W. \& Hills, J. G. 1992, ApJ, 401, 246
%\bibitem[]{410} Deloye, C. J. \& Bildsten, L. 2003, ApJ, 598, 1217
%\bibitem[]{411} De\,Marchi, G. \& Paresce, F. 1994, ApJ, 422, 597
\bibitem[]{} Deutsch, E.~W., Margon, B., \& Anderson, S.~F.\ 2000, \apjl, 530, L21 
\bibitem[]{}Dieball A., Knigge C., Zurek D.R., Shara M.M., Long K.S., Charles P.A., Hannikainen D.C., van Zyl L., 2005, ApJ, 634, L105
\bibitem[]{} Dillon, M., Gansicke, B.T., Augwerojwit, A., Rodriguez-Gil, P., Marsh, T.R., Barros, S.C.C., Szkody, P., Brady, S., Krajci, T. \& Oksanen, A. 2008, MNRAS, 386, 1568
%\bibitem[]{412} Downes, R. A., Anderson, S. F. \& Margon, B. 1996,
%PASP, 108, 688
%\bibitem[]{414} Eggleton, P. P. 1983, ApJ, 268, 368
\bibitem[]{1975MNRAS.172P..15F} Fabian A.~C., Pringle J.~E., Rees M.~J., 1975, MNRAS, 172, 15P 
%\bibitem[]{415} Guhathakurta, P., Yanni, B., Schneider, D. P. \&
%Bahcall, J. N. 1996, AJ, 111, 267
%\bibitem[]{} Hannikainen, D. C., Charles, P. A., van Zyl, L. et
%al. 2005, MNRAS, 357, 325
\bibitem[]{416} Harris, W.E. 1996, AJ, 112, 1487
%\bibitem[]{417} Haswell, C. A., Patterson, J., Thorstensen, J. R.,
%Hellier, C. \& Skillman, D. R. 1997, ApJ, 467, 847
\bibitem[]{} Hills J.G., 1976, MNRAS, 175P, 1
\bibitem[]{418} Homer, L., Charles, P. A., Naylor, T. et al. 1996, MNRAS, 282, L37
\bibitem[]{}Homer L., Anderson S.F., Margon B., Deutsch E.W., Downes R.A., 2001, ApJ, 550, L155
\bibitem[]{419} Homer, L. 2003, KITP Workshop: ``The Physics of Ultracompact Stellar Binaries'' (Feb 1-2, 2003), Coordinators: L. Bildsten, D. Chakrabarty, G. Nelemans, http://online.itp.ucsb.edu/online/ultra\_c03/homer\/
\bibitem[]{} in't Zand J.~J.~M., Jonker P.~G., Markwardt C.~B., 2007, A\&A, 465, 953 
\bibitem[]{421} Ivanova, N., Rasio, F. A., Lombardi, J. C., Dooley,  K. L. \& Proulx, Z. F. 2005, ApJ, 621, L109 
\bibitem[]{422} Katz, J. I. 1975, Nature, 253, 698
\bibitem[Nelemans(2009)]{2009arXiv0901.1778N} Nelemans, G.\ 2009, arXiv:0901.1778 
\bibitem[]{} Nelemans G., Jonker P.~G., 2006, astro, arXiv:astro-ph/0605722 
%\bibitem[]{}Nelemans G., Yungelson L.R., Portegies Zwart S.F., 2001, A\&A, 375, 890
\bibitem[]{423} Rasio, F. A., Pfahl, E. D. \& Rappaport S. 2000, ApJ, 532, L47
%\bibitem[]{424} Sirianni, M. et al.\ 2005, PASP, 117, 1049
\bibitem[]{427} Stella, L., Priedhorsky, W., \& White, N. E. 1987,  ApJ, 312, L17
\bibitem[]{} Southworth, J., Gansicke, B.T., Marsh, T.R., de Martino, D., Hakala, P., Littlefair, S., Rodriguez-Gil, P. \& Szkody, P. 2006, MNRAS, 373, 687
\bibitem[]{} Southworth, J., Marsh, T.R., Gansicke, B.T., Augwerojwit, A., Hakala, P., de Martino, D. \& Lehto, H. 2007, MNRAS, 382, 1145
\bibitem[]{} Southworth, J., Gansicke, B.T., Marsh, T.R., Torres, M.A..P., Steeghs, D., Hakala, P., Copperwheat, C.M., Aungwerojwit, A., \& Mukadam, A. 2008, MNRAS, 391, 591
%\bibitem[]{428} Stetson, P. B.\ 1991, in 3rd ESO/ST-ECF Garching -
%Data Analysis Workshop, eds. Grosb{\o}l P. J., Warmels R. H., p. 187
\bibitem[]{429} Verbunt, F. 1987, ApJ, 312, L23
\bibitem[]{} Verbunt F., 2005, in ``Interacting Binaries: Accretion, Evolution, and Outcomes, AIP Conference Proceedings, AIP Conference Series, 797, 30  
\bibitem[]{}Verbunt F., Lewin W.H.G., 2006, in {\it Compact Stellar X-ray Sources}, Cambridge University Press: Cambridge, UK (W.Lewin, M. van der Klis eds.)

%\bibitem[]{430} White, N. E. \& Angelini, L. 2001, ApJ, 561, L101
\end{thebibliography}
\end{document}